\documentstyle[12pt,epsfig]{article}
\def\be{\begin{equation}} \def\ee{\end{equation}} \def\bea{\begin{eqnarray}}
\def\eea{\end{eqnarray}} \def\nnb{\nonumber}

\begin{document}

\hfill{November 21, 2017}
%\hfill{November 21, 2017,\ \ \ {\tt bsa12C5p}}
%\hfill{July 19, 2017}

\begin{center}
\vskip 6mm 
\noindent
{\large\bf  
Elastic $\alpha$-$^{12}$C scattering at low energies 
with the bound states of $^{16}$O 
in effective field theory
}
\vskip 6mm 

\noindent
{\large 
Shung-Ichi Ando\footnote{mailto:sando@sunmoon.ac.kr}, 
}
\vskip 6mm
\noindent
{\it
School of Mechanical and ICT convergence engineering,
Sunmoon University,
Asan, Chungnam 31460, Republic of Korea
}
\end{center}

\vskip 2mm

The elastic $\alpha$-$^{12}$C scattering 
for $l=0,1,2,3$ channels at low energies
is studied, including the energies of excited bound states of $^{16}$O,
in effective field theory.
We introduce a new renormalization method 
due to the large suppression factor produced by the Coulomb interaction
when fitting the effective range parameters to the phase sift data. 
After fitting the parameters, 
we calculate asymptotic normalization constants 
of the $0_2^+$, $1_1^-$, $2_1^+$, $3_1^-$ states of $^{16}$O.
We also discuss the uncertainties of the present study when the amplitudes
are interpolated to the stellar energy region 
of the $^{12}$C($\alpha$,$\gamma$)$^{16}$O reaction.

\vskip 6mm
\noindent
PACS: 
11.10.Ef, % Lagrangian and Hamiltonian approach
24.10.-i, % Nuclear reaction models and methods
25.55.Ci, % Elastic and inelastic scattering
26.20.Fj  % Stellar helium burning

\newpage
\vskip 2mm \noindent
{\bf 1. Introduction}

The radiative $\alpha$ capture on carbon-12, 
$^{12}C(\alpha,\gamma)^{16}$O, is one of the fundamental 
reactions in nuclear-astrophysics, which determines the $C/O$ ratio 
synthesized in the stars~\cite{f-rmp84}.
The reaction rate of the process at the Gamow peak energy,
$T_G=0.3$~MeV, however, cannot be determined in experiment 
due to the Coulomb barrier. 
It is necessary to employ a theoretical model to extrapolate 
the reaction rate down to $T_G$ by fitting model parameters
to experimental data typically measured at a few MeV. 
During a last half century, a lot of experimental and theoretical
studies for the reaction have been carried out. 
See 
Refs.~\cite{bb-npa06,chk-epja15,bk-ppnp16,detal-17} for review. 

The elastic $\alpha$-$^{12}$C scattering at low energies 
is an important reaction
to fix some parameters of a model for the study.
Accurate measurements of the elastic scattering 
have been reported in Refs.~\cite{petal-npa87,tetal-prc09},
and those data provide indispensable input for the parameter fittings.
Elastic scattering data at low energies in general can be used for
deducing an asymptotic normalization constant (ANC),
which determines an overall strength of a nuclear reaction 
involving bound states~\cite{mt-prc99,by-pan08,tetal-rpp14}.  

The ANC of deuteron, for example, where the deuteron is a simple system 
consisting of loosely bound proton and neutron, leads to an overall factor
of the reactions at low energies, such as radiative neutron capture on a proton 
at BBN energies~\cite{r-npa00,aetal-prc06} 
and proton-proton fusion 
in the Sun~\cite{kr-npa99,bc-plb01,aetal-plb08,cetal-plb13}. 
The ANC of deuteron is accurately determined by two effective range 
parameters: the deuteron binding momentum and effective 
range~\cite{bs-npa01,ah-prc05}, 
which are accurately fixed from 
the deuteron binding energy and elastic $NN$ scattering 
at low energies.
On the other hand,
to deduce ANCs for nuclear reactions relevant in nuclear-astrophysics
is not so simple: 
the Coulomb interaction between heavier nuclei plays 
a negative role by preventing ones from obtaining elastic scattering 
data at very low energies, which makes the deduction of ANCs 
in terms of effective range expansion 
difficult~\cite{klh-jpg13,oin-prc16}.
Recently, a new method of the parameterization 
for deducting the ANCs of nuclear reactions
is suggested by Ramirez Suarez 
and Sparenberg~\cite{rss-16},
and new results of the ANCs by using the new method 
are reported in Refs.~\cite{bkms-prc17,oin-17}.    

Effective field theories (EFTs) 
provide us a model independent and systematic method
for theoretical calculations.
An EFT for a system in question can be built by introducing
a scale which separates relevant degrees of freedom at low energies
from irrelevant degrees of freedom at high energies.
An effective Lagrangian is written down in terms of the relevant degrees
of freedom and perturbatively expanded
by counting the number of derivatives order by order.
The irrelevant degrees of freedom are integrated out
and their effect is embedded in coefficients appearing in the Lagrangian.
Thus, a transition amplitude is systematically
calculated by writing down Feynman diagrams,
while the coefficients appearing in the Lagrangian
are fixed by experiment.
For review, one may refer to 
Refs.~\cite{bv-arnps02,bh-pr06,m-15,hjp-17}.
For last two decades, various processes essential in nuclear-astrophysics
have been investigated
by constructing EFTs, which are
$p(n,\gamma)d$ at BBN energies~\cite{r-npa00,aetal-prc06}
and $pp$ fusion~\cite{kr-npa99,bc-plb01,aetal-plb08,cetal-plb13},
$^3$He($\alpha,\gamma)^7$Be~\cite{hrv-16} and 
${}^7$Be($p$,$\gamma$)${}^7$B~\cite{znp-prc14,rfhp-epja14}
in the Sun.

In our previous work~\cite{a-epja16}, 
we have constructed an EFT of the radiative capture reaction, 
$^{12}$C($\alpha$,$\gamma$)$^{16}$O,
obtained the counting rules for the reaction at $T_G$,
and fitted some parameters of the theory to the phase shift data
of the elastic scattering. 
(We briefly review the counting rules
for the radiative capture and elastic scattering reactions 
in the following sections.) 
In the parameter fitting to the phase shift data, 
we have introduced resonance energies of $^{16}$O as 
a large scale of the theory. 
As suggested by Teichmann~\cite{t-pr51}, 
below the resonance energies, the Breit-Wigner-type parameterization 
for resonances can be expanded in powers of the energy, and
one can obtain an expression of the amplitude in terms of 
the effective range expansion.
We have determined three effective range parameters of the 
elastic scattering  for $l=0,1,2$ channels 
by fitting them to the phase sift data, 
but not included the excited bound states of $^{16}$O
in the study.
Though the phase shift data below the resonance energies 
can be reproduced very well 
by using the fitted parameters, 
we find that significant uncertainties 
in the elastic amplitudes are remained
when extrapolating them to $T_G$.
 
In the present work, we incorporate the excited binding energies 
for $0_2^+$, $1_1^-$, $2_1^+$, $3_1^-$ ($l^\pi_{i-th}$) states of $^{16}$O 
in the parameter fitting to the phase shift data of the elastic scattering
for $l=0,1,2,3$ channels.
Our assumption for the parameter fitting is 
that fitted curves which interpolate the amplitude between 
the phase shift data and the excited binding energies 
can be represented by several terms of a polynomial function.
As will be discussed in detail below, however,
we find a mismatch between the strength of the amplitudes estimated
from the phase shift data and the first few terms 
of a polynomial function obtained from the Coulomb self-energy term
in the dressed $^{16}$O propagator. 
Because those terms from the Coulomb self-energy are larger, 
at most by two order of magnitude, 
than the term estimated by the phase shift data, 
we introduce a new renormalization method;
we assume that those large terms should be renormalized 
by counter terms, the role of which 
we assign to the effective range terms. 
Thus
we include the effective range parameters up to third order 
($n=3$ in powers of $k^{2n}$) 
for the $l=0,1,2$ channels and up to fourth order 
($n=4$) for the $l=3$ channel.
After fitting the parameters to the phase shift data, 
we calculate the ANCs of the 
$0_2^+$, $1_1^-$, $2_1^+$, $3_1^-$ 
states of $^{16}$O and compare our results to the existing ones. 

This paper is organized as the following:
In Sec. 2, the approach based on an EFT for the radiative capture reaction
is briefly reviewed, and the expression of equations 
related to the elastic scattering 
amplitudes, the phase shifts, 
and the effective range parameters are displayed.
In Sec. 3, we introduce a new renormalization method 
and describe the details of the numerical fitting to the elastic scattering
data.
In Sec. 4, the numerical results obtained in this work are 
exhibited, and finally in Sec. 5, the results 
and discussion of the work are presented.
In Appendix, the structure of the UV divergence and the counter terms
of the elastic scattering amplitudes in the conventional renormalization 
method are summarized.

\vskip 2mm \noindent
{\bf 2. EFT for the radiative capture 
and elastic scattering at low energies}

In the study of the radiative capture process,
$^{12}$C($\alpha$,$\gamma$)$^{16}$O, at $T_G=0.3$~MeV
employing an EFT, at such a low energy, 
we regard the ground states of $\alpha$ and $^{12}$C as point-like particles 
whereas the first excited state energies of $\alpha$ and $^{12}$C 
are chosen as irrelevant degrees of freedom, by which 
a large scale of the theory is determined. 
The effective Lagrangian for the process is constructed 
in terms of two spinless scalar fields for $\alpha$ and $^{12}$C,
and the terms of the Lagrangian are expanded in terms of 
the number of derivatives. 
An expression of the effective Lagrangian has been obtained in
Eq.~(1) in Ref.~\cite{a-epja16}.
The expansion parameter of the theory 
is $Q/\Lambda_H \sim 1/3$ where $Q$ denotes a typical
momentum scale $Q\sim k_G$: $k_G$ is the Gamow peak momentum,
$k_G = \sqrt{2\mu T_G}\simeq 41$~MeV, where
$\mu$ is the reduced mass of $\alpha$ and $^{12}$C. 
$\Lambda_H$ denotes a large momentum scale
$\Lambda_H\simeq \sqrt{2\mu_4 T_{(4)}}$ or 
$\sqrt{2\mu_{12} T_{(12)}}\sim 150$~MeV where 
$\mu_4$ is the reduced mass of
one and three-nucleon system and $\mu_{12}$ is that of four and 
eight-nucleon system. $T_{(4)}$ and $T_{(12)}$ are 
the first excited energies 
of $\alpha$ and $^{12}$C, respectively.
By including the terms up to next-to-next-to-leading order,
therefore,
one may obtain about 10\% theoretical uncertainty
for the process. 

\begin{figure}[t]
\begin{center}
\includegraphics[width=14cm]{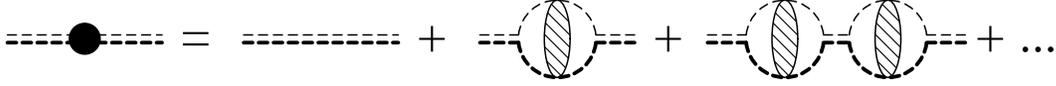}
\caption{Diagrams for dressed $^{16}$O propagator.
A thick (thin) dashed line represents a propagator of $^{12}$C ($\alpha$), 
and a thick and thin double dashed line with and without a filled blob
represent a dressed and bare $^{16}$O propagator, respectively.
A shaded blob represents 
a set of diagrams consisting of all possible one-potential-photon-exchange 
diagrams up to infinite order and no potential-photon-exchage one.
%the off-shell Coulomb $T$-matrix.
}
\label{fig;dressed-propagator}
\end{center}
\end{figure}
\begin{figure}[t]
\begin{center}
\includegraphics[width=3.5cm]{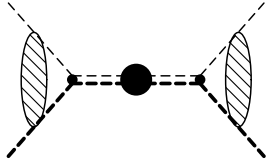}
\caption{Diagram of the scattering amplitude. 
See the caption of Fig.~1 as well.
}
\label{fig;scattering-amplitude}
\end{center}
\end{figure}
The amplitudes of the elastic scattering 
are calculated from diagrams depicted in 
Figs.~\ref{fig;dressed-propagator} and \ref{fig;scattering-amplitude}.
In our previous works, 
we have obtained the scattering amplitudes for $l$-th partial wave states
as~\cite{a-epja16,aetal-prc07,a-epja07} 
\bea
A_l &=& \frac{2\pi}{\mu}\frac{(2l+1)P_l(\cos\theta) e^{2i\sigma_l}
W_l(\eta)
C_\eta^2}{
K_l(k) - 2\kappa H_l(k)
}\,,
\label{eq;Al}
\eea
where $k$ is the magnitude of relative momentum between $\alpha$ and $^{12}$C
and $\theta$ is the scattering angle in the C.M. frame. 
In addition,
$\eta$ is the Sommerfeld parameter, $\eta = \kappa/k$, 
where $\kappa$ is the inverse of the Bohr radius, 
$\kappa = Z_2Z_6\mu \alpha$, and % = 244.760$~MeV, and
\bea
C_\eta^2 = \frac{2\pi\eta}{e^{2\pi\eta}-1}\,,
\ \ \ 
W_l(\eta) = \frac{\kappa^{2l}}{(l!)^2}
\prod^{l}_{n=0}\left(
1+\frac{n^2}{\eta^2}
\right)\,,
\ \ \ 
H_l(k) = W_l(\eta) H(\eta)\,,
\eea
with
\bea
H(\eta) = \psi(i\eta) + \frac{1}{2i\eta} -\ln(i\eta)\,.
\label{eq;H}
\eea
$\psi(z)$ is the digamma function,
$P_l(x)$ are the Legendre polynomials,
and $\sigma_l$ are the Coulomb phase shifts.
When $\eta$ goes to zero, the factor $C_\eta^2$ is normalized to one 
whereas, when $\eta$ becomes large, the Gamow factor, 
$P=\exp(-2\pi\eta)$, appears from the factor $C_\eta^2\propto  P$.
We note that the function, $-2\kappa H_l(k)$, in the denominator
of the amplitude is obtained from the Coulomb bubble diagram 
for the dressed propagator of $^{16}$O
in Fig.~\ref{fig;dressed-propagator}, and the factor,
$e^{2i\sigma_l}W_l(\eta)C_\eta^2$, in the numerator 
is from the initial and final state Coulomb interactions
between $\alpha$ and $^{12}$C
in Fig.~\ref{fig;scattering-amplitude}.  

The function $K_l(k)$ represents the interaction
due to the short range nuclear force 
(compared with the long range Coulomb force), 
which is obtained
in terms of the effective range parameters as~\footnote{
In this work, we employ a modified representation
for effective range parameters from that presented 
in Ref.~\cite{scb-jp11}.
Here we use the effective volume-like parameter $P_l$ rather than
the shape parameter $P_l$ 
represented as $-r_l^2P_lk^4$. 
In addition, we introduced an opposite sign 
for the $R_l$ term so as to have positive sign 
in the bounding energy in Eq.~(\ref{eq;moneoveral}). 
We had employed another parameterization
($v$ parameterization) for the effective range parameters 
in Ref.~\cite{ah-prc12}.
}   
\bea
K_l(k) = -\frac{1}{a_l} + \frac12 r_l k^2 - \frac14 P_l k^4
+ Q_l k^6 - R_l k^8 
+ \cdots\,.
\label{eq;Kl}
\eea
Because the UV divergence comes out of the loop integrals in the 
Coulomb self-energy terms (from the diagrams 
in Fig.~\ref{fig;dressed-propagator}), we need to introduce 
counter terms for renormalization. We employ the dimensional regularization
and the structure of the UV divergence from the Coulomb self-energy 
terms is given in Appendix. Thus, we need to include one counter term,
$-1/a_0$ term in the effective range expansion for $l=0$,
two counter terms, $-1/a_1$ and $r_1$ for $l=1$, 
three counter terms, $-1/a_2$, $r_2$, and $P_2$ for $l=2$, and
four counter terms, $-1/a_3$, $r_3$, $P_3$, and $Q_3$ for $l=3$, 
to remove the UV divergence and make the terms finite. 
The other higher order terms in the effective range expansion
are introduced as finite terms in the conventional renormalization method
and supposed to obey counting rules in which higher order terms are 
less important than lower order terms.
We find that the expression obtained in Eq.~(\ref{eq;Al}) 
reproduces well the previous results reported
in Refs.~\cite{klh-jpg13,hot-npb73,h-jmp77,scb-prc10}.

At the binding energies of excited states of $^{16}$O,
the amplitudes should have a pole 
at $k_b = i\gamma_l$ where $\gamma_l$ are the binding momenta~\footnote{
The quantity $\gamma_l$ is also referred to the bound-state wave number 
in the low energy scattering theory.
}, 
$\gamma_l = \sqrt{2\mu B_l}$;
$B_l$ denote the binding energies of excited states of $^{16}$O.
Thus the denominator of the scattering amplitude, $D_l(k)$,
should vanish at $k_b$;
\bea
D_l(k_b) = K_l(k_b) - 2\kappa H_l(k_b) = 0\,.
\label{eq;flb}
\eea
Using this condition, the first effective range parameter, $a_l$,
is related to other effective range parameters as
\bea
-\frac{1}{a_l} &=& 
\frac12 r_l\gamma_l^2 
+ \frac14 P_l\gamma_l^4
+ Q_l \gamma_l^6
+ R_l \gamma_l^8
+ \cdots
+2\kappa H_l(k_b)\,,
\label{eq;moneoveral}
\eea 
and we remove the $a_l$ dependence from the amplitude.
Thus, we have
$D_l(k)$ as 
\bea
D_l(k) &=& \frac12r_l\left(k^2+\gamma_l^2\right)
- \frac14P_l\left(k^4-\gamma_l^4\right)
+ Q_l\left(k^6 + \gamma_l^6\right)
-R_l\left(k^8 -\gamma_l^8\right)
+ \cdots
\nnb \\ && 
-2\kappa\left[H_l(k)-H_l(k_b)\right]\,.
\eea
The remaining effective range parameters are fixed by using the phase
shift data of the elastic scattering.

The differential cross section of the elastic scattering 
is represented in terms of the pure Coulomb scattering part
and the Coulomb modified nuclear scattering part, as presented
in Eq.~(3) in Ref.~\cite{a-epja16}, where
the scattering function $U_l$ is 
$U_l=\exp[2i(\delta_l+\omega_l)]$ where $\delta_l$ are the phase 
shifts for $l$-th partial waves and $\omega_l=\sigma_l-\sigma_0$
with $\sigma_l=\arg \Gamma(1+l+i\eta)$.
Thus the scattering amplitudes are represented in terms of 
$\delta_l$ as~\cite{l-rmp57}
\bea
A_l &=& \frac{2\pi}{\mu}\frac{(2l+1)P_l(\cos\theta)e^{2i\sigma_l}}{
k\cot \delta_l-ik}\,. 
\label{eq;Al-delta}
\eea
By comparing two expressions of the amplitudes $A_l$ 
in Eqs.~(\ref{eq;Al}) and (\ref{eq;Al-delta}), 
one has a relation between the phase shift 
and the effective range parameters in $D_l(k)$ as 
\bea
W_l(\eta)C_\eta^2 k\cot\delta_l = Re D_l(k)\,.
\label{eq;dell}
\eea

To estimate the ANC, $|C_b|$, 
for the $0_2^+$, $1_1^-$, $2_1^+$, $3_1^-$ states of $^{16}$O, 
we employ the definition of $|C_b|$
from Eq.~(14) in Ref.~\cite{scb-prc10}:
\bea
|C_b| &=& 
\gamma_l^l\frac{\Gamma(l+1+|\eta_b|)} {l!}
\left(\left|-\frac{dD_l(k)}{dk^2}\right|_{k^2=-\gamma_l^2}\right)^{-\frac12}\,
\ \ \ \mbox{\rm (fm$^{-1/2}$)}\,,
\label{eq;Cb}
\eea 
where $\eta_b=\kappa/k_b$.

\vskip 2mm \noindent
{\bf 3. Fitting the parameters to phase shifts data }

Four excited states of $^{16}$O 
exist below the $\alpha$-$^{12}$C threshold, 
which we include in the parameter fitting
in the present study. 
The binding energies, $B_i(l^\pi)$, 
of the $i$-th excited bound states of $^{16}$O in $l^\pi$ states
from the $\alpha$-$^{12}$C threshold energy are
$B_1(0^+) = 1.113$, 
$B_2(3^-) = 1.032$, 
$B_3(2^+) = 0.245$, 
$B_4(1^-) = 0.045$~MeV.
Thus, the binding momenta, $\gamma_l=\sqrt{2\mu B_i(l^\pi)}$, are
\bea
\gamma_l = 
79.843,\
15.860,\
37.007, \ 
75.954\,\ \ \mbox{\rm (MeV)}\,,
\eea
for the $0_2^+$, $1_1^-$, $2_1^+$, $3_1^-$ states, respectively,
where $\mu = m_\alpha m_C/(m_\alpha + m_C) = 2795.079\,$MeV with
$m_\alpha = 3727.379$~MeV and $m_C = 11174.862$~MeV.
As mentioned above, the first effective range term, $a_l$, 
is constrained by using the binding momenta. 

To fix the other effective range parameters, 
the phase shift data for each $l$-th partial wave state 
are used. 
In the present work, we employ the phase shit data 
from the Tischhauser {\it et al.}'s paper~\cite{tetal-prc09}. 
The reported energies of the $\alpha$ particle in the lab. frame
are $T_\alpha=2.6$-6.6~MeV, 
and corresponding momenta in the C.M. frame are $k=105$-166~MeV
(i.e., $k= \sqrt{1.5\mu T_\alpha}$).
Because our large momentum scale
of the theory is $\Lambda_H\sim 150$~MeV, 
the convergence of the expansion series should be carefully
examined when the elastic scattering data are used for the parameter fitting. 
In addition, because we do not explicitly include the resonance
states of $^{16}$O in the theory, we restrict data sets for the parameter
fitting
below the resonance energies, $T_\alpha=6.52$, 3.23, 3.57, 5.09~MeV 
for $0_3^+$, $1_2^-$, $2_2^+$, $3_2^-$ states, respectively;
the corresponding momenta are $k=166$, 117, 123, 146~MeV for
the $l=0$, 1, 2, 3, channels, respectively. 
(We will mention data sets we choose 
for the parameter fitting in detail below.)       

We are now in position to discuss a new renormalization method. 
The effective range parameters in $K_l(k)$ 
are expanded in powers of $k^2$ whereas 
the real part of the function $H_l(k)$
can be expanded in powers of $k^2$ as well. 
For the function $H(\eta)$ in $H_l(k)$, one has
\bea
Re H(\eta) &=& 
\frac{1}{12\kappa^2}k^2 
+ \frac{1}{120\kappa^4}k^4
+ \frac{1}{252\kappa^6}k^6
+ \frac{1}{240\kappa^8}k^8
+ \cdots\,,
\eea  
where $\eta = \kappa/k$; $\kappa$ is the inverse of the Bohr radius, 
$\kappa\simeq 245$~MeV, and is regarded as another large scale of the theory.
This expansion is reliable in our study for the elastic scattering 
at low energies along with the effective range expansion in $K_l(k)$.
Thus, the right-hand-side of equation, $Re D_l(k)$, in Eq.~(\ref{eq;dell})
can be expanded as a power series of $k^2$ for 
both $K_l(k)$ and $2\kappa Re H_l(k)$.
Meanwhile, the left-hand-side 
of Eq.~(\ref{eq;dell}) is suppressed by the factor $C_\eta^2$,
due to the Gamow factor $P=\exp(-2\pi\eta)$.

In the case of the $s$-wave, for example, the reported phase shift 
at the smallest energy, $T_\alpha = 2.6$~MeV, 
is $\delta_0=-1.893^\circ$~\cite{tetal-prc09}. 
The factor $C_\eta^2$ becomes 
$C_\eta^2\simeq 6 \times 10^{-6}$ at $k=104$~MeV
which corresponds to $T_\alpha=2.6$~MeV, and 
the left-hand-side of Eq.~(\ref{eq;dell}) numerically becomes   
$C_\eta^2 k\cot\delta_0 = -0.019$~MeV. 
The function $2\kappa Re H_0(k)$ is expanded as
\bea
2\kappa Re H_0(k) &=& 
\frac{1}{6\kappa}k^2 
+ \frac{1}{60\kappa^3}k^4
+ \frac{1}{126\kappa^5}k^6
+ \frac{1}{120\kappa^7}k^8 + \cdots
\nnb \\ &=& 7.441 + 0.136 + 0.012 + 0.002 + \cdots \ \ 
\mbox{\rm (MeV)}\,,
\label{eq;S-wave}
\eea
at $k=104$~MeV. The numerical values in the second line 
of Eq.~(\ref{eq;S-wave}) correspond to the terms appearing 
in the first line of the equation in order. 
One can see that the power series converges well, but
the first and second terms are two and one order of magnitude 
larger compared with the value estimated by using the experimental data 
in the left-hand-side of Eq.~(\ref{eq;dell}),
$-$0.019~MeV.
Thus, we regard those terms unnaturally large, and it is  
necessary to introduce a new renormalization method, 
in which the counter terms remove the unnaturally large terms 
and make the terms in a natural size.
In other words,
we assume that fitting polynomial functions 
are represented as a natural power series at the low energy region,
and to maintain such polynomial functions, 
large cancellations for the first and second terms 
with the $r_l$ and $P_l$ effective terms, respectively, are expected.
So we include the three effective range parameters,
$r_l$, $P_l$, and $Q_l$, for the $l=0$ channel, 
as the counter terms. 
The same tendency can be seen in the $l=1,2$ channels whereas 
one needs four effective range parameters for the $l=3$ channel.
Thus, we employ the three effective range parameters, 
$r_l$, $P_l$, $Q_l$ for the $l=0,1,2$ channels 
and the four effective range parameters,  
$r_l$, $P_l$, $Q_l$, $R_l$  for the $l=3$ channel 
when fitting the parameters to the phase shift data below.\footnote{
In the new method of the parameterization of elastic scattering 
suggested by Ramirez Suarez and Sparenberg, the $K_l(k)$ 
and $2\kappa H_l(k)$ functions are merged, and a new
function for the parameterization 
is defined as $\Delta_l(E)=C_\eta^2 k \cot\delta_l$, 
which is parameterized 
by using the Pade approximation~\cite{rss-16}.
}

\vskip 2mm \noindent
{\bf 4. Numerical results}

As discussed above, we fit the three effective range parameters,
$r_l$, $P_l$ and $Q_l$ to the phase shift data 
for the $l=0,1,2$ channels
and the four effective range parameters, $r_l$, $P_l$, $Q_l$, and $R_l$ 
to those for the $l=3$ channel, 
while $a_l$ are constrained by using the relation 
in Eq.~(\ref{eq;moneoveral}) 
with the binding momenta $\gamma_l$.
To examine the sensitivity to the choice of data sets, 
we employ three sets of the phase shift data~\cite{tetal-prc09}
below the resonance energy for each partial wave, 
which have different energy ranges: 
three data sets for $l=0$ denoted by $S0$, $S1$, $S2$ 
have the data at $T_\alpha = 2.6$-3.6, 2.6-3.8, 2.6-4.0~MeV, respectively,
those for $l=1 (2)$ denoted by $P0$, $P1$, $P2$ ($D0$, $D1$, $D2$) 
have the data at $T_\alpha = 2.6$-3.0, 2.6-3.1, 2.6-3.2~MeV, respectively,  
and those for $l=3$ denoted by $F0$, $F1$, $F2$ have the data at
$T_\alpha = 2.6$-4.6, 2.6-4.8, 2.6-5.0~MeV, respectively.  

When the parameters are fitted to the data,
large cancellations between the terms in powers of $k^2$ 
appearing from the $K_l(k)$ and $2\kappa H_l(k)$ functions,
the $r_l$, $P_l$, $Q_l$, $R_l$ effective range terms and 
from the $2\kappa H_l(k)$ function, are expected. 
We denote the terms from the $2\kappa H_l(k)$ function 
corresponding to the effective range terms as $\tilde{r}_l$,
$\tilde{P}_l$, $\tilde{Q}_l$, $\tilde{R}_l$, and we have  
\bea
&&
\tilde{r}_0 =  \frac{1}{3\kappa}\,,
\ \ \
\tilde{P}_0 = - \frac{1}{15\kappa^3}\,,
\ \ \
\tilde{Q}_0 = \frac{1}{126\kappa^5}\,,
\label{eq;tilde_values_s-wave}
\\ &&
\tilde{r}_1 =  \frac{1}{3}\kappa\,,
\ \ \
\tilde{P}_1 = - \frac{11}{15\kappa}\,,
\ \ \
\tilde{Q}_1 = \frac{31}{1260\kappa^3}\,,
\label{eq;tilde_values_p-wave}
\\ &&
\tilde{r}_2 =  \frac{1}{12}\kappa^3\,,
\ \ \
\tilde{P}_2 = - \frac{51}{60}\kappa\,,
\ \ \
\tilde{Q}_2 = \frac{191}{1008\kappa}\,,
\label{eq;tilde_values_d-wave}
\\ &&
\tilde{r}_3 =  \frac{1}{108}\kappa^5\,,
\ \ \
\tilde{P}_3 = - \frac{47}{180}\kappa^3\,,
\ \ \
\tilde{Q}_3 = \frac{5297}{22680}\kappa\,,
\ \ \ 
\tilde{R}_3 = - \frac{17101}{90720\kappa}\,.
\label{eq;tilde_values_f-wave}
\eea
Those values (and variations from them for some terms) 
are used as the initial input of the effective range parameters 
for the parameter fitting.\footnote{
We employ a SciPy module, {\tt curve\_fit}, in optimization package
when fitting the effective range parameters to the phase shift data. 
} 

\begin{table}
\begin{center}
\begin{tabular}{c|cccc|cc} \hline
     & $a_0$ (fm) & $r_0$ (fm) & $P_0$ (fm$^3$) & $Q_0$ (fm$^5$) &
$ReD_{0G}$ (MeV) & $|C_b|$ (fm$^{-1/2}$) \cr \hline
$S0$ & $6.2\times 10^4$ & 0.268514(3) & $-0.0343(4)$ & 0.0019(2) &
$4.2(7)\times 10^{-3}$ & $6.8(16)\times 10^2$ \cr
$S1$ & $6.6\times 10^4$ & 0.268514(3) & $-0.0342(3)$ & 0.0020(3) &
$4.0(5)\times 10^{-3}$ & $7.4(15)\times 10^2$ \cr
$S2$ & $5.8\times 10^4$ & 0.268513(3) & $-0.0345(2)$ & 0.0018(1) &
$4.4(4)\times 10^{-3}$ & $6.4(7)\times 10^2$ \cr \hline
 & --- & $\tilde{r}_0$ (fm) & $\tilde{P}_0$ (fm$^3$) & $\tilde{Q}_0$ (fm$^5$) &
--- & --- \cr \hline
 &  ---   & 0.268735    & $-0.0349$    & 0.0027 & ---  & ---
\cr \hline 
\end{tabular}
\caption{
Effective range parameters, $r_0$, $P_0$, $Q_0$, fitted 
by using the data sets, $S0$, $S1$, $S2$;
values of $\tilde{r}_0$, $\tilde{P}_0$, $\tilde{Q}_0$ 
are included in last raw. 
The values of $a_0$, $ReD_{0G}$, and $|C_b|$ for the $0_2^+$ state 
are calculated by using $r_0$, $P_0$, $Q_0$. 
For details, see the text.
}
\label{table:s-wave_parameters}
\end{center}
\end{table}
\begin{table}
\begin{center}
\begin{tabular}{c|cccc|cc} \hline
 &$a_1$(fm$^3$)& $r_1$ (fm$^{-1}$) & $P_1$ (fm) & $Q_1$ (fm$^3$) &
$ReD_{1G}$ (MeV$^3$) & $|C_b|$ (fm$^{-1/2}$) \cr \hline
$P0$ & $-1.8\times 10^5$ & 0.4150(6) & $-0.577(8)$ & 0.019(3) &
$2.7(8)\times 10^2$ & $1.9(4)\times 10^{14}$ \cr
$P1$ & $-1.6\times 10^5$ & 0.4153(2) & $-0.574(2)$ & 0.020(1) &
$3.0(3)\times 10^2$ & $1.8(1)\times 10^{14}$ \cr
$P2$ & $-1.3\times 10^5$ & 0.4157(2) & $-0.569(2)$ & 0.023(1) &
$3.5(3)\times 10^2$ & $1.6(1)\times 10^{14}$ \cr \hline 
&---&$\tilde{r}_1$ (fm$^{-1}$) & $\tilde{P}_1$ (fm) & $\tilde{Q}_1$ (fm$^3$) &
 --- & --- \cr \hline
 & ---  & 0.4135& $-0.591$ & 0.013 & --- & --- \cr \hline 
\end{tabular}
\caption{
Effective range parameters, $r_1$, $P_1$, $Q_1$, fitted 
by using the data sets, $P0$, $P1$, $P2$;
values of $\tilde{r}_1$, $\tilde{P}_1$, $\tilde{Q}_1$ 
are included in last raw.   
The values of $a_1$, $ReD_{1G}$, and $|C_b|$ for the $1_1^-$ state 
are calculated by using $r_1$, $P_1$, $Q_1$.
For details, see the text.
}
\label{table:p-wave_parameters}
\end{center}
\end{table}
\begin{table}
\begin{center}
\begin{tabular}{c|cccc|cc} \hline
 &$a_2$(fm$^5$)& $r_2$ (fm$^{-3}$) & $P_2$ (fm$^{-1}$) & $Q_2$ (fm) &
$ReD_{2G}$ (fm$^{-5}$) & $|C_b|$ (fm$^{-1/2}$) \cr\hline
$D0$ & $10.3\times 10^3$ & 0.155(4) & $-1.12(7)$ & 0.11(3) &
$-1.66(156)\times 10^{-4}$ & $2.4(3)\times 10^4$ \cr
$D1$ & $6.5\times 10^3$ & 0.152(2) & $-1.16(4)$ & 0.08(2) &
$-2.6(9)\times 10^{-4}$ & $2.3(2)\times 10^4$ \cr
$D2$ & $4.3\times 10^3$ & 0.149(2) & $-1.21(3)$ & 0.06(1) &
$-3.8(6)\times 10^{-4}$ & $2.1(1)\times 10^4$ \cr \hline
 & --- & $\tilde{r}_2$ (fm$^{-3}$) & $\tilde{P}_2$ (fm$^{-1}$) & 
$\tilde{Q}_2$ (fm) & --- & --- \cr \hline
 & --- & 0.159& $-1.05$ & 0.15 & --- & --- \cr \hline 
\end{tabular}
\caption{
Effective range parameters, $r_2$, $P_2$, $Q_2$, fitted 
by using the data sets, $D0$, $D1$, $D2$;
values of $\tilde{r}_2$, $\tilde{P}_2$, $\tilde{Q}_2$ 
are included in the last raw.  
The values of $a_2$, $ReD_{2G}$, and $|C_b|$ for the $2_1^+$ state 
are calculated by using $r_2$, $P_2$, $Q_2$.
For details, see the text.
}
\label{table:d-wave_parameters}
\end{center}
\end{table}
\begin{table}
\begin{center}
\begin{tabular}{c|ccccc|cc} \hline
 &$a_3$(fm$^7$)&$r_3$(fm$^{-5}$)&$P_3$(fm$^{-3}$)&$Q_3$(fm$^{-1}$)&$R_3$(fm)&
$ReD_{3G}$ (fm$^{-7}$) & $|C_b|$ (fm$^{-1/2}$) \cr \hline
$F0$ & $-1.4\times 10^3$ & 0.0319(1) & $-0.453(11)$ & 0.317(9)&$-0.141(8)$&
$7.8(8)\times 10^{-4}$ & $1.2(1)\times 10^2$ \cr
$F1$ & $-1.5\times 10^3$ & 0.0320(1) & $-0.459(9)$ & 0.311(7)&$-0.146(6)$&
$7.4(7)\times 10^{-4}$ & $1.3(1)\times 10^2$ \cr
$F2$ & $-1.8\times 10^3$ & 0.0322(1) & $-0.472(7)$ & 0.301(6)&$-0.156(5)$&
$6.4(6)\times 10^{-4}$ & $1.5(1)\times 10^2$ \cr \hline
 & --- & $\tilde{r}_3$ (fm$^{-5}$) & $\tilde{P}_3$ (fm$^{-3}$) & 
$\tilde{Q}_3$ (fm$^{-1}$) & $\tilde{R}_3$ (fm) & --- & --- \cr \hline
 & --- & 0.0272& $-0.498$ & 0.290 &$-0.152$ & --- & --- \cr\hline 
\end{tabular}
\caption{
Effective range parameters, $r_3$, $P_3$, $Q_3$, $R_3$, fitted 
by using the data sets, $F0$, $F1$, $F2$;
values of $\tilde{r}_3$, $\tilde{P}_3$, $\tilde{Q}_3$, $\tilde{R}_3$ 
are included in the last raw.
The values of $a_3$, $ReD_{3G}$, and $|C_b|$ for the $3_1^-$ state 
are calculated by using $r_3$, $P_3$, 
$Q_3$, $R_3$.  For details, see the text.
}
\label{table:f-wave_parameters}
\end{center}
\end{table}
In Tables 
\ref{table:s-wave_parameters}, 
\ref{table:p-wave_parameters}, 
\ref{table:d-wave_parameters}, 
\ref{table:f-wave_parameters}, 
fitted values and errors of the effective range parameters,
$r_l$, $P_l$, $Q_l$, $R_l$ for $l=0,1,2,3$ channels 
to the data sets 
$\{S0, S1, S2\}$, 
$\{P0, P1, P2\}$, 
$\{D0, D1, D2\}$, 
$\{F0, F1, F2\}$, respectively, 
are presented. 
The errors of the fitted parameters stem 
from those of the phase shift data. 
Numerical values of the $\tilde{r}_l$, $\tilde{P}_l$, $\tilde{Q}_l$,
$\tilde{R}_l$ terms for $l=0,1,2,3$ 
are also shown in the tables.
The values of $a_l$ are calculated by using the fitted
effective range parameters in Eq.~(\ref{eq;moneoveral}). 
The values in the second last column 
are the real part of the denominator $ReD_l(k)$ 
of the scattering amplitude at the energy corresponding to 
$T_G$~\footnote{
The $\alpha$ energy in the lab. frame corresponding to $T_G$ is
$T_\alpha = \frac43 T_G \simeq 0.4$~MeV.
}
(i.e., at $k=k_G$),
and those in the last column 
are the ANC, $|C_b|$,
for the $0_2^+$, $1_1^-$, $2_1^+$, $3_1^-$ states. 

One can see the errors of the fitted values of the effective range
parameters are small, but the fitted effective range parameters are 
largely canceled with the corresponding $\tilde{r}_l$, $\tilde{P}_l$,
$\tilde{Q}_l$, $\tilde{R}_l$ terms.
In the values of the real part of denominator, $ReD_{lG}$, 
of the scattering amplitude at $k=k_G$, we find significant errors: 
about 9-17\%, 9-30\%, 16-94\%, 9-10\% errors, depending on the 
choice of the data sets, for the $l=0,1,2,3$ channels, respectively.
A large uncertainty persists in the $l=2$ channel. 

The ANCs for the $1_1^-$ and $2_1^+$ states have intensively been 
studied because those ANCs are related to the estimate of the $E1$ and $E2$ 
transitions of the radiative capture process, while those for the 
$0_2^+$ and $3_1^-$ states, which are also important to estimate
the cascade transitions, are recently studied in experiment
and reported first time. 

For the ANC, $|C_b|$, for the $1_1^-$ state, we find that our result, 
$|C_b|= (1.6$-$1.9)\times 10^{14}$ (fm$^{-1/2}$), is in good agreement 
with experimental values, 
$(2.10\pm 0.14) \times 10^{14}$,
$(2.00\pm 0.35) \times 10^{14}$,
$(2.08\pm 0.20) \times 10^{14}$,
obtained 
by Avila {\it et al.}~\cite{aetal-prl15},
Oulebsir {\it et al.}~\cite{oetal-prc12},
Brune {\it et al.}~\cite{betal-prl99}, respectively,
and underestimates for other experimental ones, 
$(5.1\pm0.6)\times 10^{14}$~\cite{betal-npa07} 
and $(17.4$-26.4)$\times 10^{14}$~\cite{ab-plb11}.
We also find good agreement with theoretical estimates,
$(2.22$-2.24)$\times 10^{14}$, obtained from a potential model calculation
by Katsuma~\cite{k-prc08}, 
and $2.14(6) \times 10^{14}$ and $2.073\times 10^{14}$  from 
the new method of the parameterization 
by Ramirez Suarez and Sparenberg~\cite{rss-16}
and by Orlov {\it et al.}~\cite{oin-17}, respectively.

For the ANC, $|C_b|$, for the $2_1^+$ state, our result, 
$|C_b| = (2.1$-$2.4) \times 10^4$~(fm$^{-1/2}$),
is in underestimates to experimental values, 
$(12.2\pm 0.7)\times 10^4$~\cite{aetal-prl15},
$(14.4\pm 2.8)\times 10^4$~\cite{oetal-prc12},
$(11\pm 1)\times 10^4$~\cite{betal-prl99},
$(34.5\pm 0.5)\times 10^4$~\cite{betal-npa07},
(12.2-18.2)$\times 10^4$~\cite{ab-plb11}. 
Other experimental estimates evaluated earlier, which basically 
agree with the experimental values mentioned above, can be found in 
Table VI in Ref.~\cite{dd-prc08}.
On the other hand, our result is in good agreement with theoretical 
estimates, $(2.41\pm 0.38)\times 10^4$ and $2.106\times 10^4$, 
from the effective range analysis
up to the $r_2$ term by Konig {\it et al.}~\cite{klh-jpg13} and 
up to the $P_2$ term by Orlov {\it et al.}~\cite{oin-prc16}, 
respectively, and 
in underestimates for the other theoretical estimates,
$(14.45\pm 0.85)\times 10^4$ from the supersymmetric potential model 
by Sparenberg~\cite{s-prc04} and 
($12.6\pm 0.5)\times 10^4$ from the $R$ matrix analysis
with a microscopic cluster model by Dufour and Descouvemont~\cite{dd-prc08} 
and $5.050\times 10^4$ from the new method of the parameterization
by Orlov {\it et al.}~\cite{oin-17}.

For the ANCs, $|C_b|$, for the $0_2^+$ and $3_1^-$ states, 
our result, $|C_b| = (6.4$-$7.4)\times 10^2$ (fm$^{-1/2}$) for the
$0_2^+$ state is in underestimate to an experimental value,
$(15.6\pm 1.0) \times  10^2$~\cite{aetal-prl15} and in overestimate to 
a theoretical value, $4.057\times 10^2$~\cite{oin-17}.
Meanwhile, our result, $|C_b|=(1.2$-$1.5)\times 10^2$ (fm$^{-1/2}$)
for the $3_1^-$ state
is in very good agreement to the experimental value, 
$(1.39\pm 0.09)\times 10^2$, recently reported 
by Avila {\it et al.}~\cite{aetal-prl15}. 

\begin{table}
\begin{center}
\begin{tabular}{c|cccc} \hline
 & $|-\frac{1}{a_l}|$ & $|\frac12(r_l-\tilde{r}_l)k_G^2|$ & 
 $|-\frac14(P_l-\tilde{P}_l)k_G^4|$ & $|(Q_l-\tilde{Q}_l)k_G^6|$ \cr \hline
$S2$ & 1 & 0.276 & 0.012 & 0.004 \cr
$P2$ & 0.154 & 1 & 0.215 & 0.016 \cr
$D2$ & 1 & 0.946 & 0.316 & 0.031 \cr
$F2$ & 1 & 0.195 & 0.023 & 0.002 \cr \hline
\end{tabular}
\caption{
Ratios of the terms in the power series 
to $-1/a_l$ for $l=0,2,3$ and to $\frac12(r_1-\tilde{r}_1)k_G^2$ 
for $l=1$ at $k=k_G$ where 
the effective range parameters fitted by using the data sets,
$S2$, $P2$, $D2$, $F2$, have been used. 
}
\label{table;ratios}
\end{center}
\end{table}
To examine the convergence of the power series in terms of $k^2$ at $k=k_G$, 
we add the effective range terms and those 
from the $2\kappa ReH_l(k)$ functions together. 
In Table \ref{table;ratios}, we show the ratios of the terms
after normalizing those terms by $-1/a_l$ for $l=0,2,3$ and
by $\frac12(r_1-\tilde{r}_1)k_G^2$ for $l=1$ because of their dominance
where the effective range parameters fitted by using the data sets,
$S2$, $P2$, $D2$, $F2$, are used. 
As discussed above, the expansion parameter at $T_G$ is $Q\sim 1/3$, 
so the $k_G^2$, $k_G^4$, $k_G^6$ terms are expected to be  
a few tenth, a few hundredth, a few thousandth to the leading order terms, 
respectively.
We find good convergence of the power series for $l=0,3$ at $k=k_G$.
On the other hand, the $-1/a_1$ term is small compared to 
the $\frac12(r_1-\tilde{r}_1)k_G^2$ term for $l=1$ and the $-1/a_2$ and 
$\frac12(r_2-\tilde{r}_2)k_G^2$ terms are comparable for $l=2$, 
but the higher order terms are well converged for $l=1,2$ as expected 
by the counting rules of the theory.

\begin{figure}
\begin{center}
\includegraphics[width=7.5cm]{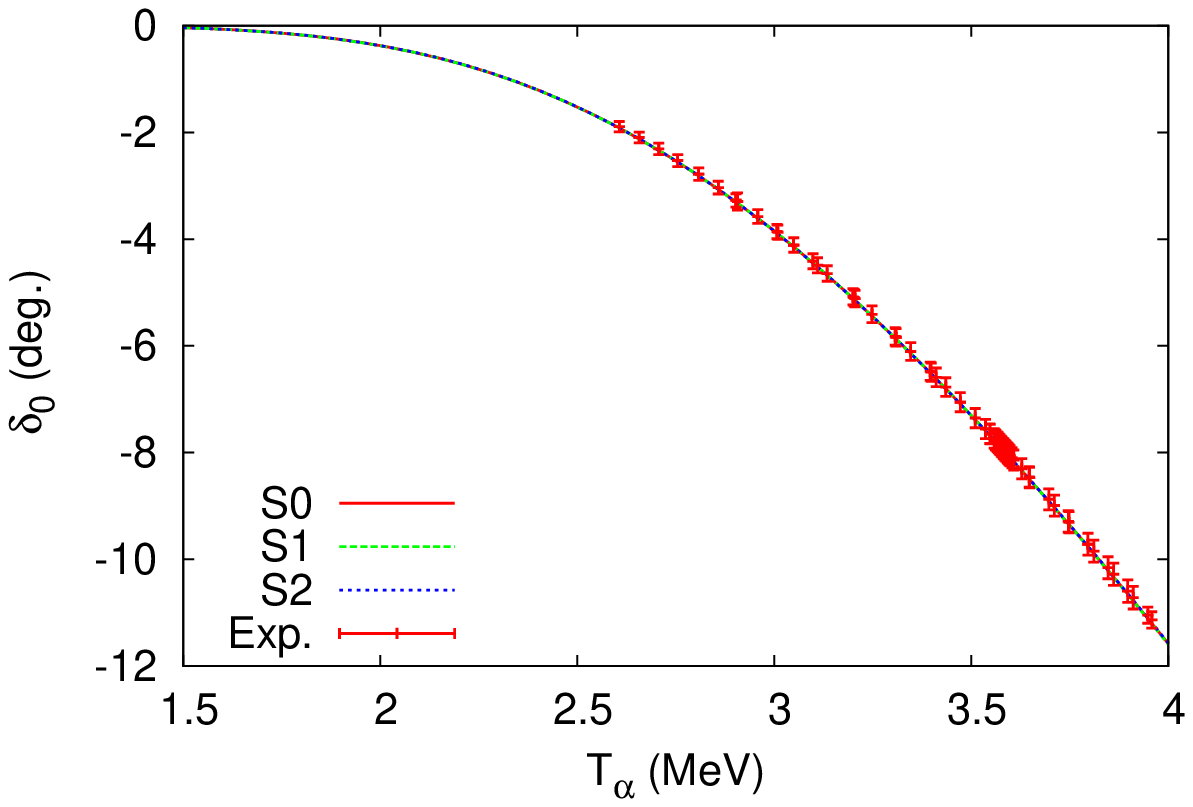}
\includegraphics[width=7.5cm]{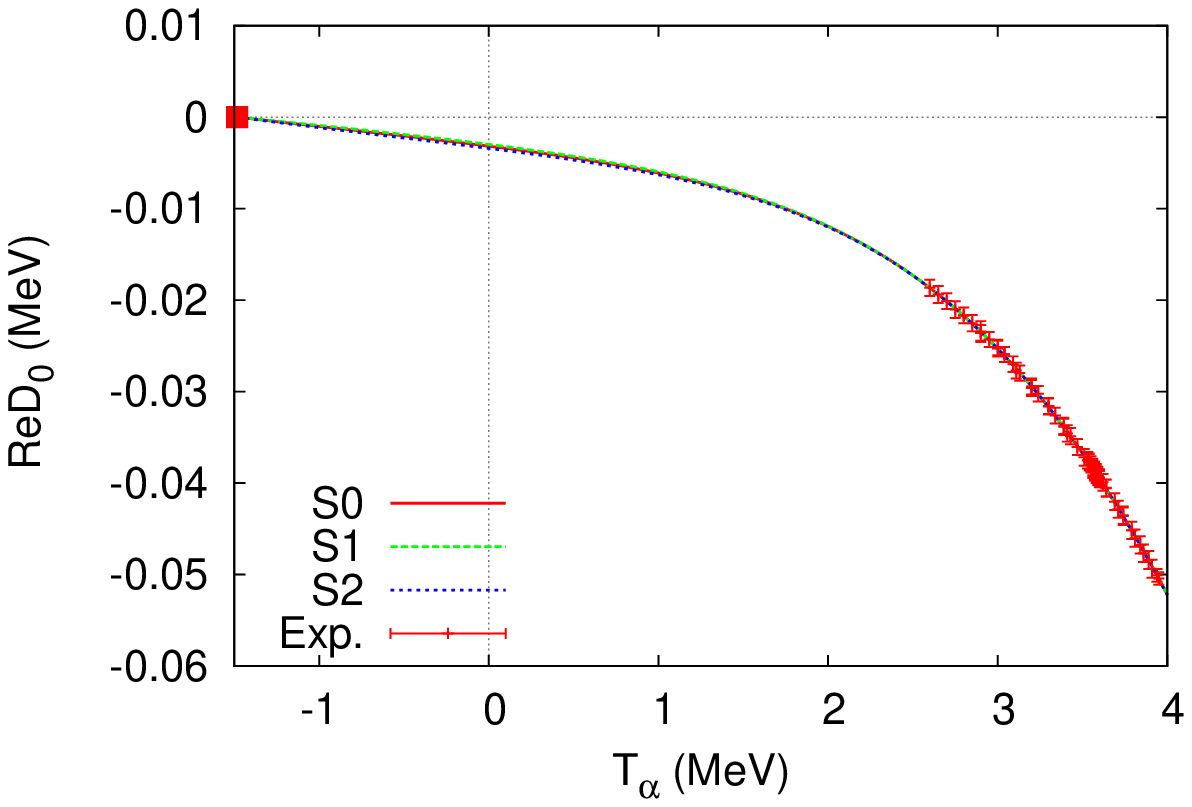}
\caption{
Phase shift, $\delta_0$, (left panel) and  
the real part of denominator, $ReD_0(k)$, of the amplitude (right panel)
as functions of $T_\alpha$ (where $k=\sqrt{1.5\mu T_\alpha}$).
Curves are plotted by using the effective range
parameters, fitted from the $S0$, $S1$, $S2$ data sets,
presented in Table \ref{table:s-wave_parameters}.
Exp. phase shift data are also included in the figure.
A filled box in the right panel represents the excited binding energy 
of the $0^+_2$ state.
}
\label{fig;sw}
\end{center}
\end{figure}

\begin{figure}
\begin{center}
\includegraphics[width=7.5cm]{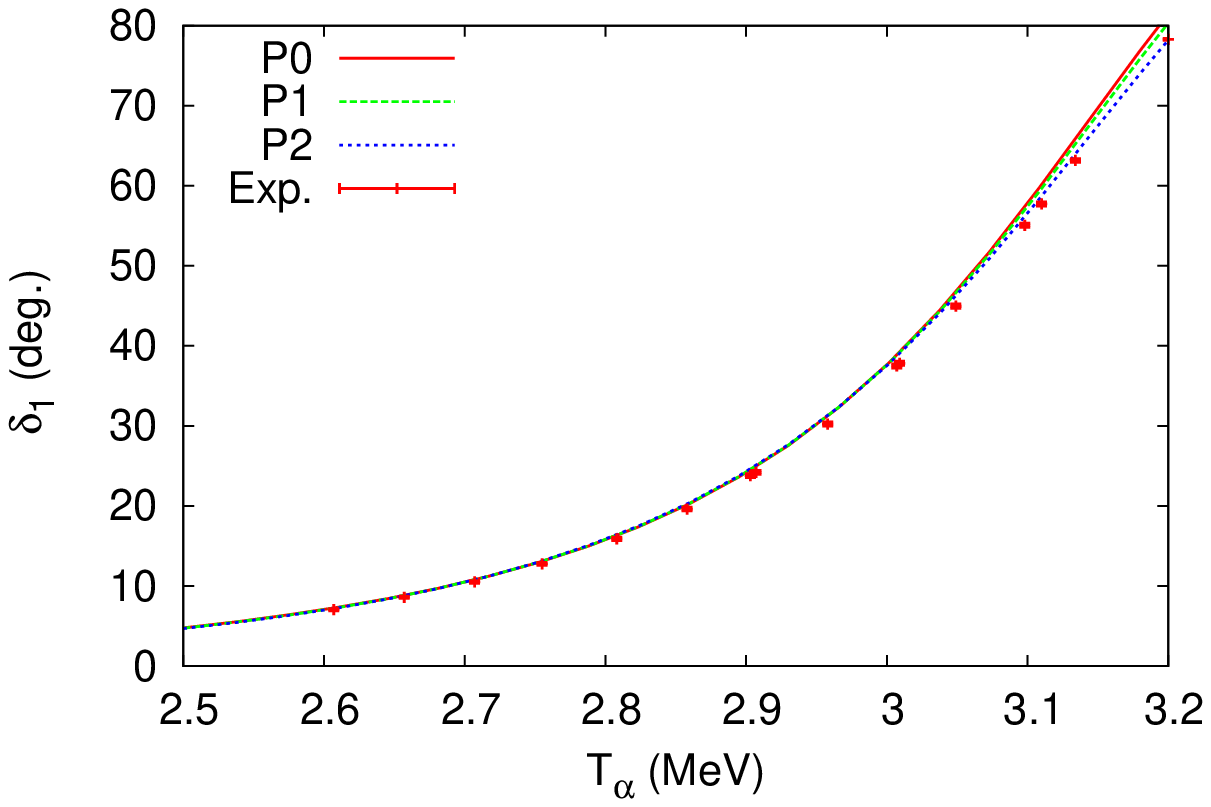}
\includegraphics[width=7.5cm]{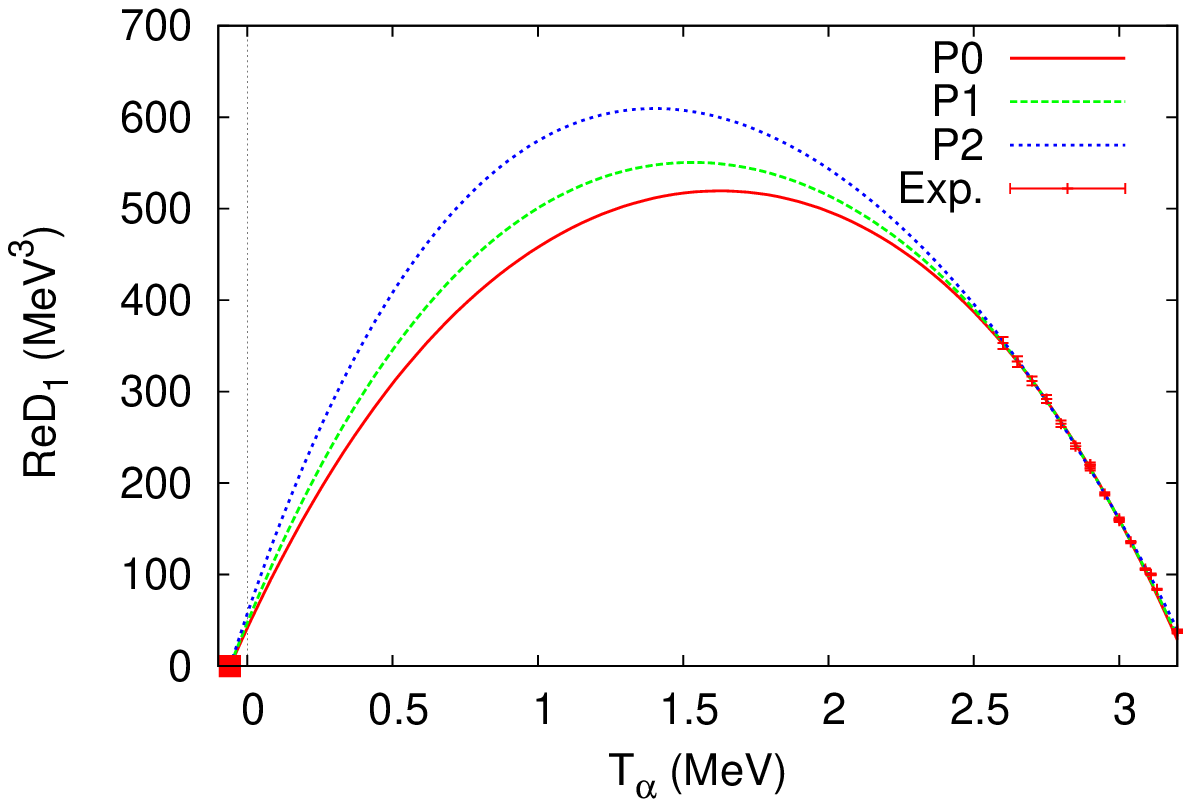}
\caption{ 
Phase shift, $\delta_1$, (left panel) and  
the real part of 
denominator, $ReD_1(k)$, of the amplitude (right panel)
as functions of $T_\alpha$.
Curves are plotted by using the effective range
parameters, fitted from the $P0$, $P1$, $P2$ data sets,
 presented in Table \ref{table:p-wave_parameters}.
Exp. phase shift data are also included in the figure.
A filled box in the right panel represents 
the binding energy of the $1^-_1$ state.
}
\label{fig;pw}
\end{center}
\end{figure}

\begin{figure}
\begin{center}
\includegraphics[width=7.5cm]{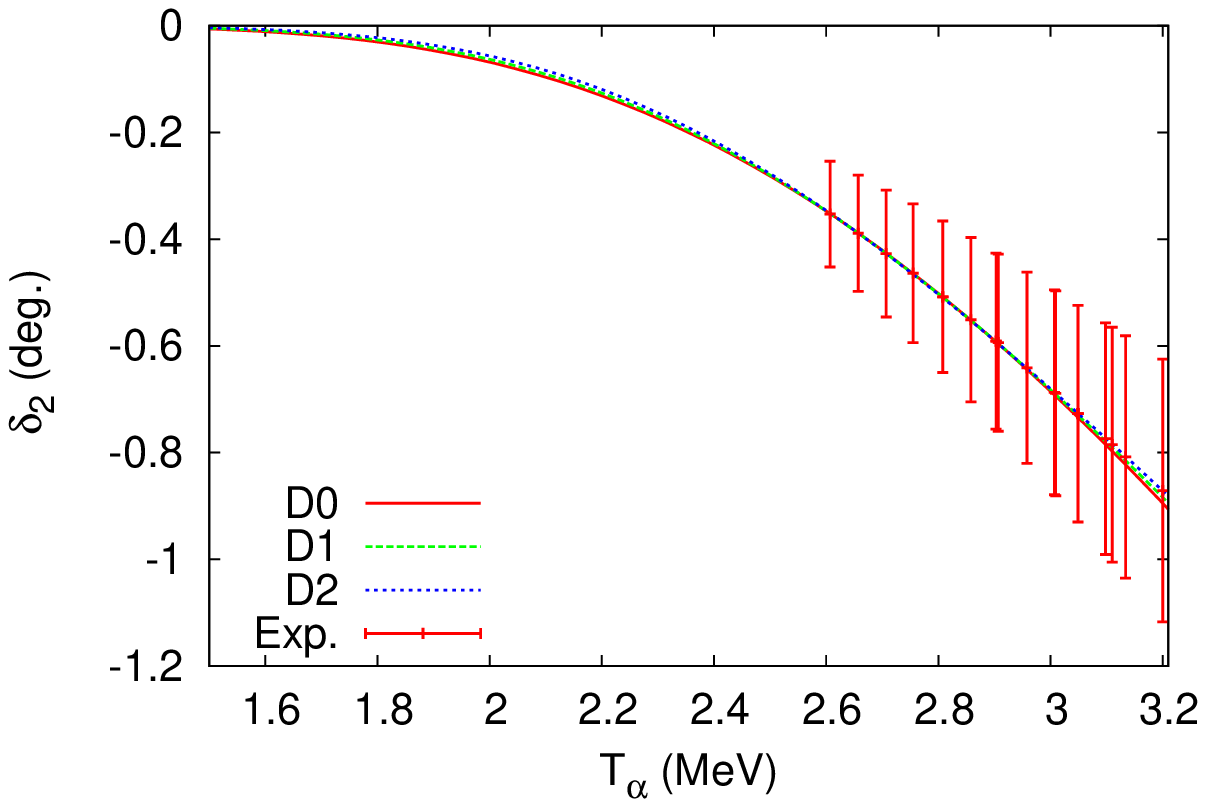}
\includegraphics[width=7.5cm]{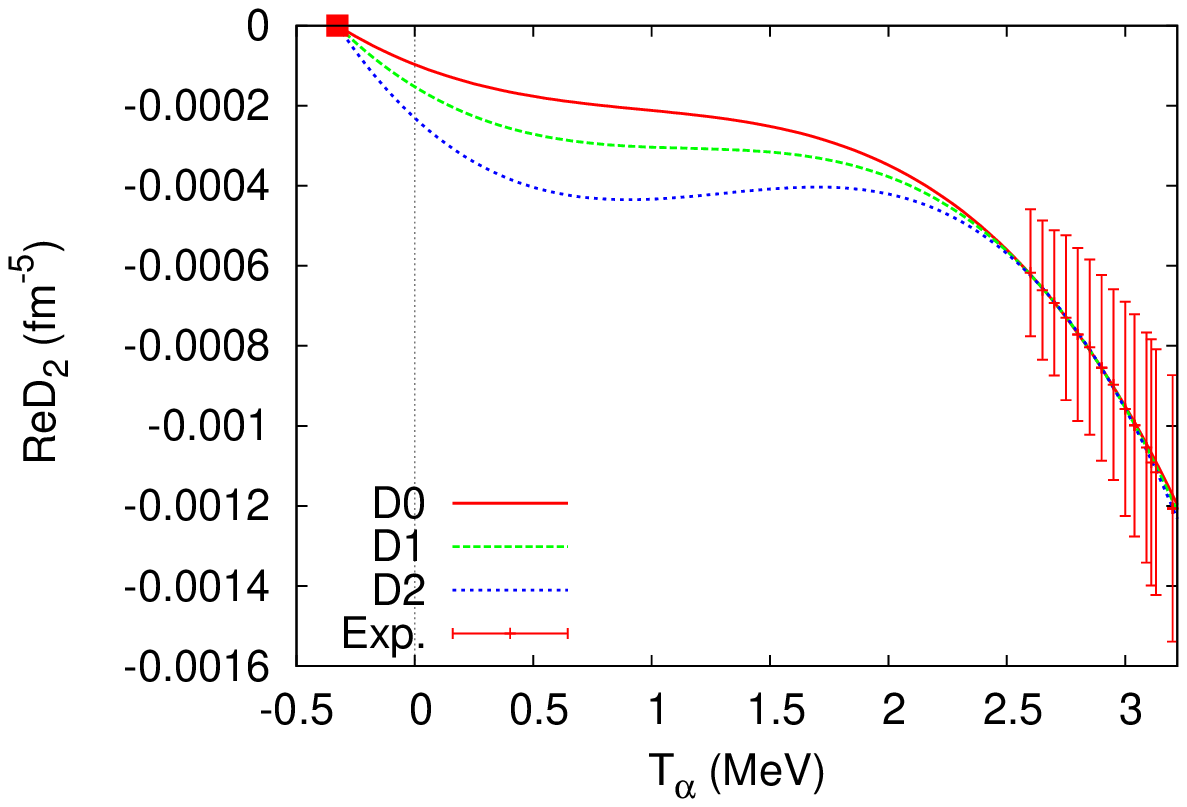}
\caption{ 
Phase shift, $\delta_2$, (left panel) and  
the real part of denominator, $ReD_2(k)$, of the amplitude (right panel)
as functions of $T_\alpha$.
Curves are plotted by using the effective range
parameters, fitted from the $D0$, $D1$, $D2$ data sets,
presented in Table \ref{table:d-wave_parameters}.
Exp. phase shift data are also included in the figure.
A filled box in the right panel represents the binding energy 
of the $2^+_1$ state.
}
\label{fig;dw}
\end{center}
\end{figure}

\begin{figure}
\begin{center}
\includegraphics[width=7.5cm]{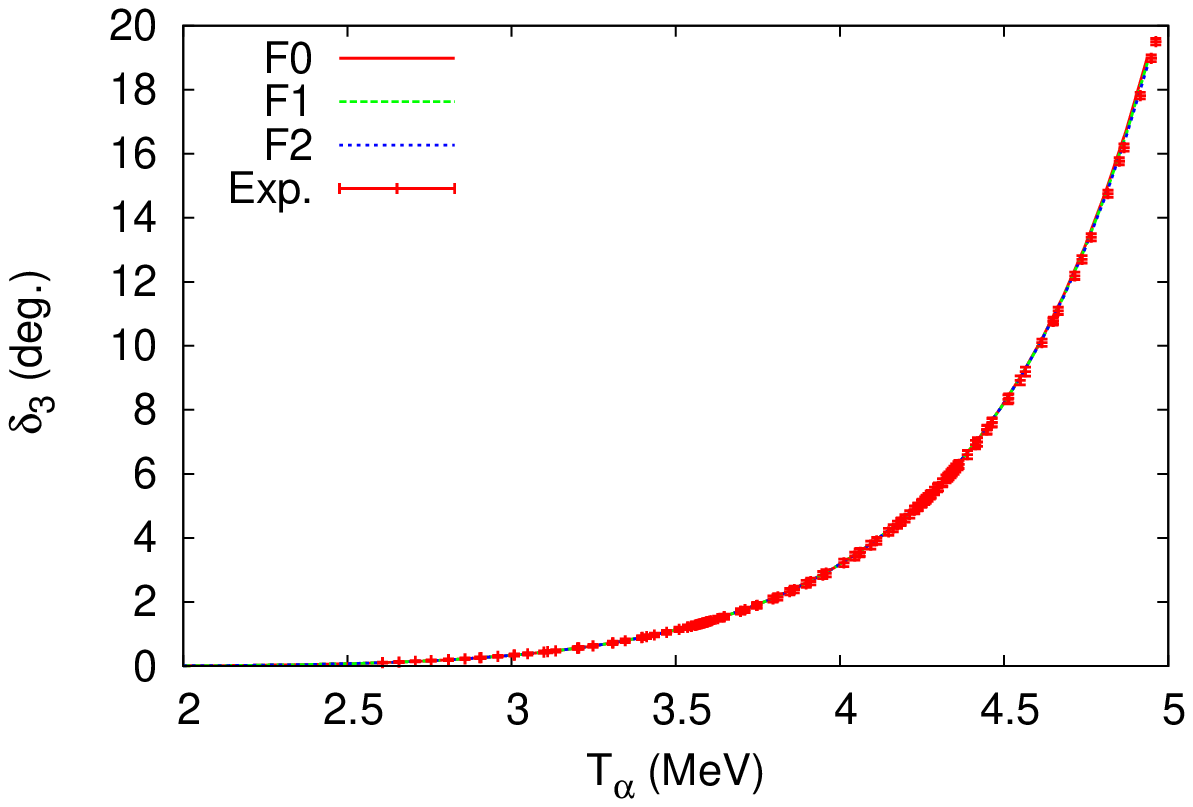}
\includegraphics[width=7.5cm]{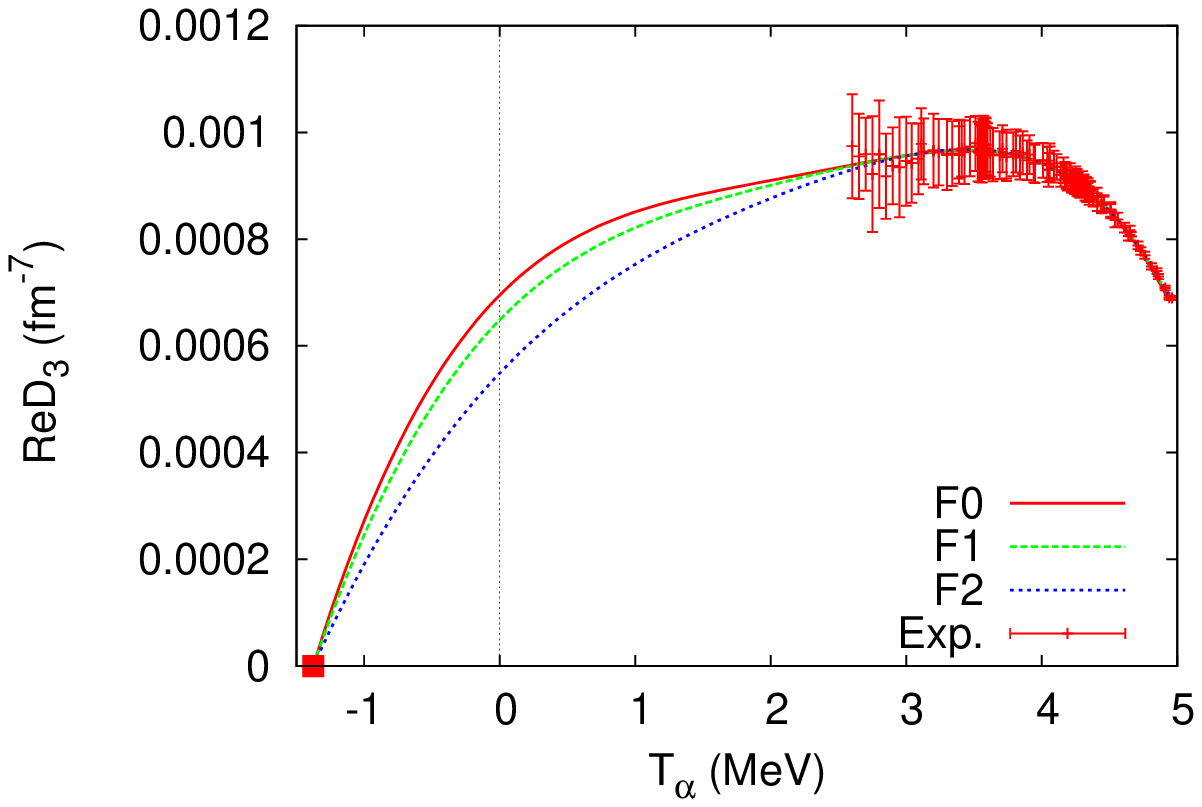}
\caption{ 
Phase shift, $\delta_3$, (left panel) and  
the real part of denominator, $ReD_3(k)$, of the amplitude (right panel)
as functions of $T_\alpha$.
Curves are plotted by using the effective range
parameters, fitted from the $F0$, $F1$ $F2$ data sets,
presented in Table \ref{table:f-wave_parameters}.
Exp. phase shift data are also included in the figure.
A filled box in the right panel represents the binding energy 
of the $3^-_1$ state.
}
\label{fig;fw}
\end{center}
\end{figure}

In Figs.~\ref{fig;sw}, \ref{fig;pw}, \ref{fig;dw}, \ref{fig;fw}, 
the curves of phase shift $\delta_l$ (left panels)
and the real part of denominator, $ReD_l(k)$, 
of the scattering amplitude (right panels)
for $l=0,1,2,3$, respectively, 
are plotted as functions of $T_\alpha$ 
by using the values of $r_l$, $P_l$, $Q_l$, $R_l$,
which are fitted by using the data sets denoted by \{$S0, S1, S2$\},
\{$P0,P1,P2$\}, \{$D0,D1,D2$\}, \{$F0,F1,F2$\}.
Experimental data of the phase shift are also included in the figures.
In addition, a filled box in the right panel represents the binding
energy of the excited $0^+_2$, $1_1^-$, $2_1^+$, or $3_1^-$ state in each
of the figures. 

We find that the curves of $\delta_l$ plotted by using 
the different sets of the fitted parameters are in good agreement 
with each other and reproduce the experimental data within the errors, 
except for the large energy region, $T_\alpha=
3.0$-3.2~MeV for $l=1$.
One can see the significant separations
of the curves of $ReD_l(k)$ at the interpolated energy region 
where the experimental data do not exist for the $l=1,2,3$ channels, 
but the values of $ReD_{lG}$ at $k=k_G$ 
(i.e., $T_\alpha = 0.4$~MeV) for the different sets 
of the parameters are still in good agreement within the errors,
as we have seen in the tables.   

\vskip 2mm \noindent
{\bf 5. Results and discussion}

In the present work, we have fitted the effective range parameters
to the phase shift data of the elastic scattering for $l=0,1,2,3$ 
below the resonance energies of $^{16}$O in EFT.
The excited binding energies of the $0_2^+$, $1_1^-$, $2_1^+$,
$3_1^-$ states of $^{16}$O are also included in the parameter fitting.
Because of a mismatch between the terms from 
the $2\kappa H_l(k)$ functions and a term obtained from the 
phase shift data, we have introduced a new renormalization method: 
we assign the effective range terms
as a role of the counter terms so as to obtain a natural power series 
for the fitting polynomial functions
at the low energy region. Thus, we have fitted three effective range
parameters, $r_l$, $P_l$, $Q_l$, for $l=0,1,2$ and four effective 
range parameters, $r_l$, $P_l$, $Q_l$, $R_l$, for $l=3$ to the 
phase shift data.   
(Those fitted values of the effective range parameters are used 
when we study the radiative capture reaction of $\alpha$ and $^{12}$C 
in EFT in the future.)
After fitting the effective range parameters,
we have calculated the real part of the denominator, $ReD_{lG}$, of the 
scattering amplitude at the energy corresponding to $T_G$
and the ANCs for the $0_2^+$, $1^-_1$, $2^+_1$, $3_1^-$ states.  
In addition, 
we have interpolated and plotted 
the real part of the denominator of the scattering amplitude
between the binding energy and the phase shift data.
  
In fitting the effective range parameters for the all partial waves,
we find that the errors of the fitted effective range parameters 
are tiny whereas the effective range terms are almost exactly canceled
with the terms from the $2\kappa H_l(k)$ function. Thus, we obtain
9-94\% errors in $ReD_{lG}$ depending on the choice of the input 
data sets and partial waves, 
while the power series in terms of $k_G^2$ in $ReD_l(k)$ 
well converges at the energy corresponding to $T_G$, as expected by
the counting rules of the theory.
In the figures, though the curves of the phase shifts plotted 
by using the different sets of the parameters are 
in good agreement, those of $ReD_l(k)$ are significantly 
different in the interpolated region between the binding energies 
and the phase shift data where no experimental data are available.
Nonetheless, $ReD_{lG}$ from the different sets of the parameters
are still in good agreement within the error bars. 

For the ANC, $|C_b|$, for the $1_1^-$ state, we find 
our result is in good agreement with the other theoretical estimates and 
the recent experimental estimates.
Thus, the estimates of $|C_b|$ for the $1_1^-$ state 
converge both theoretically and experimentally.
For the ANC, $|C_b|$, for the $2_1^+$ state, our result is in good agreement
with the theoretical estimates based on the effective range expansion
but in underestimates, more than by the factor of 5, 
compared with those of the other theories and the experiments.
As seen in Eq.~(\ref{eq;Cb}),
such a large $|C_b|$ can be obtained by a very small slope of $ReD_2(k)$ 
at the binding energy of the $2_1^+$ state.
That indicates a very large scattering length and a drastic cancelation
between the $r_2$ term and the $\tilde{r}_2$ term. 
Meanwhile, as seen in the right panel of Fig.~\ref{fig;dw},
the phase shift data are quite distant from the bound state energy,
while the higher order terms involve in the fitting.
Thus, it is hard to discriminate which curve is better than the others 
in the present approach.
To have accurate experimental data of the phase shift 
down to, e.g., $T_\alpha=1$ or 1.5~MeV could improve the situation.  
For the ANCs, $|C_b|$, for the $0_2^-$ and $3_1^-$ states, 
the first experimental result is recently reported 
by Avila {\it et al.}~\cite{aetal-prl15}. We find that
our result for the $0_2^+$ state is about a half compared to the 
experimental estimate, while our result for the $3_1^-$ state is in 
good agreement to the experimental value. It may be necessary 
to wait for a further confirmation experimentally and theoretically 
for the $0_2^+$ and $3_1^-$ states.

In the present work, we have introduced a new renormalization method
from an observation of a mismatch between the terms from the Coulomb
self-energy term, the $2\kappa H_l(k)$ function, and the term obtained 
from the phase shift data, by assuming to have a natural power series
of the fitting polynomial functions at the low energies.  
Our conjecture about the observation is, on one hand, that
it may be caused simply due to the severe suppression factor, 
the Gamow factor, at the low energies. 
On the other hand, it may stem from our assumption that
the $\alpha$ and $^{12}$C states are point-like. 
That implies that the interaction length scale between 
the $\alpha$ and $^{12}$C vanishes, and thus the short range
effect should be renormalized by introducing the counter terms. 
A more systematic study about the issue, indeed, would be necessary 
in the future.

\vskip 2mm \noindent
{\bf Acknowledgements}

The author would like to thank J.-M. Sparenberg for communications.
This work was supported by
the Basic Science Research Program through the National Research
Foundation of Korea funded by the Ministry of Education of Korea
(Grant No. NRF-2016R1D1A1B03930122)
and in part by
the National Research Foundation of Korea (NRF)
grant funded by the Korean government
(Grant No. NRF-2016K1A3A7A09005580).

\vskip 2mm \noindent
{\bf Appendix }

In this appendix, we display the UV divergent terms from 
the Coulomb self-energy in terms of the $J$ functions
and discuss the counter terms in the conventional renormalization method. 
The Coulomb self-energy terms for $l=0,1,2,3$ are calculated from the 
$J$ functions defined below: 
\bea
J^{l=0}_0(p) &=& 
\int 
\frac{d^3\vec{q}}{(2\pi)^3}
\frac{d^3\vec{q}'}{(2\pi)^3}
\langle \vec{q}'|\hat{G}_C^{(+)}|\vec{q}\rangle\,,
\\
J^{l=1}_{2,i,x}(p) &=& 
\int 
\frac{d^3\vec{q}}{(2\pi)^3}
\frac{d^3\vec{q}'}{(2\pi)^3}
q'_i
\langle \vec{q}'|\hat{G}_C^{(+)}|\vec{q}\rangle
q_x \,,
\\
J^{l=2}_{4;ij,xy}(p) &=&  
\int 
\frac{d^3\vec{q}}{(2\pi)^3}
\frac{d^3\vec{q}'}{(2\pi)^3}
\left(q'_iq'_j - \frac13\delta_{ij}q'^2\right)
\langle \vec{q}'|\hat{G}_C^{(+)}|\vec{q}\rangle
\left(q_xq_y - \frac13\delta_{xy}q^2\right)
\,,
\\
J_{6,ijk,xyz}^{l=3}(p) &=& \int
\frac{d^3\vec{q}}{(2\pi)^3}
\frac{d^3\vec{q}'}{(2\pi)^3}
\left[
q'_iq'_jq'_k
- \frac15 \left(
\delta_{ij} q'_k 
+\delta_{ik} q'_j 
+\delta_{jk} q'_i 
\right) q^{'2}
\right]
\langle \vec{q}'|\hat{G}_C^{(+)}|\vec{q}\rangle
\nnb \\ && \times
\left[ q_x q_y q_z 
- \frac15 \left(
\delta_{xy} q_z 
+\delta_{xz} q_y 
+\delta_{yz} q_x 
\right) q^2
\right]\,,
\eea
where $\hat{G}_C^{(\pm)}$ is the Coulomb propagator, 
$G^{(\pm)}_C = 
1/(E - \hat{H}_0 - \hat{V}_C \pm i \epsilon)$,
with $H_0 =\vec{p}^2/(2\mu)$ is the free two particle Hamiltonian:
$\mu$ is a reduced mass, and $V_C = \alpha Z_1 Z_2/r$ 
is the repulsive Coulomb force. 

The $J$ functions become infinity due to the loop integrals, 
and we employ the dimensional 
regularization in $d=4-2\epsilon$ space-time dimensions. 
The expression of the $J_0^{l=0}(p)$ function is well known
and one has~\cite{kr-npa99}
\bea
J_0^{l=0}(p) &=& 
J_0^{l=0,div}
- \frac{\kappa\mu}{\pi}H(\eta)\,,
\eea
where $J_0^{l=0,div}$ is the divergent part of 
the $J_0^{l=0}(p)$ function,
\bea
J_0^{l=0,div} &=& 
\frac{\kappa\mu}{2\pi}\left[
\frac{1}{\epsilon} 
- 3 C_E
+ 2
+ \ln\left(
\frac{\pi\mu_{DR}^2}{4\kappa^2}
\right)
\right]\,,
\eea
where $C_E =0.577\cdots$ and $\mu_{DR}$ is a scale parameter 
from the dimensional regularization.
In addition, $\kappa = \alpha Z_1Z_2\mu$, and 
the $H(\eta)$ function has been presented in Eq.~(\ref{eq;H}).
We note that $J_0^{l=0,div}$ does not depend on the momentum $p$.
Thus the divergent term is renormalized a counter term, 
$-1/a_0$ term in the effective range expansion for $l=0$.

The UV divergent terms from the other $J$ functions are obtained as 
\bea
J^{l=1,div}_{2,i,x}(p) &=&
O_{i,x}^{l=1} (\kappa^2+p^2) J^{l=0,div}_{0}\,,
\\
J_{4,ij,xy}^{l=2,div}(p) &=& 
O_{ij,xy}^{l=2}
\frac14
(\kappa^2+p^2)(\kappa^2+4p^2) J_0^{l=0,div}\,,
\\
J_{6,ijk,xyz}^{l=3,div}(p) &=& 
O_{ijk,xyz}^{l=3} 
\frac{1}{36}
(\kappa^2+p^2)(\kappa^2+4p^2)(\kappa^2+9p^2)J_0^{l=0,div}\,,
\eea
with
\bea
O_{i,x}^{l=1} &=& \int\frac{d\Omega_{\hat{l}}}{4\pi}
\hat{l}_i\hat{l}_x
= \frac13\delta_{ix}\,,
\\
O_{ij,xy}^{l=2} &=& \int\frac{d\Omega_{\hat{l}}}{4\pi}
\left(
\hat{l}_i\hat{l}_j - \frac13\delta_{ij}
\right)
\left(
\hat{l}_x\hat{l}_y - \frac13\delta_{xy}
\right)
= \frac{1}{15}\left(
\delta_{ix}\delta_{jy} 
+\delta_{iy}\delta_{jx}
-\frac23\delta_{ij}\delta_{xy} 
\right)\,,
\\
O_{ijk,xyz}^{l=3} &=& 
\int\frac{d\Omega_{\hat{l}}}{4\pi}
\left[\hat{l}_i\hat{l}_j\hat{l}_k
- \frac15\left(
\delta_{ij}\hat{l}_k 
+\delta_{ik}\hat{l}_j 
+\delta_{jk}\hat{l}_i 
\right)
\right]
\left[\hat{l}_x\hat{l}_y\hat{l}_z
- \frac15\left(
\delta_{xy}\hat{l}_z 
+\delta_{xy}\hat{l}_y 
+\delta_{yz}\hat{l}_x 
\right)
\right]
\nnb \\ &=& 
\frac{1}{105}\left[
\delta_{ix}\delta_{jy}\delta_{kz} + \mbox{\rm 5\ terms}
- \frac25\left(
\delta_{ij}\delta_{kx}\delta_{yz} + \mbox{\rm 8\ terms}
\right)
\right]\,.
\eea
To renormalize the divergent terms from the Coulomb self-energy, 
which now depend on the powers of $p^2$, 
one needs two counter terms,
$-1/a_1$ and $r_1$ in the effective range expansion for $l=1$,
three counter terms, $-1/a_2$, $r_2$, $P_2$ for $l=2$, and
four counter terms, $-1/a_3$, $r_3$, $P_3$, $Q_3$ for $l=3$.

\end{document}